\documentstyle[version2,preprint,prb,aps]{revtex}

\newcommand{\bee}{\begin{equation}}
\newcommand{\ene}{\end{equation}}
\newcommand{\bea}{\begin{eqnarray}}
\newcommand{\eea}{\end{eqnarray}}
\newcommand{\di}{\mbox{d}}
\newcommand{\phe}{\hat \phi}
\newcommand{\jh}{\hat j}
\newcommand{\bG}{{\bf \Gamma}}
\newcommand{\vphi}{\varphi}
\newcommand{\vphe}{\hat \varphi}

\begin{document}
\draft

\begin{title}{Long Wavelength Anomalous Diffusion Mode in the 2D XY Dipole 
Magnet.} 
\end{title}

\author{Ar.Abanov$^{1}$, A.Kashuba$^{1,2,3}$ and V.L.Pokrovsky$^{1,3}$.}

\begin{instit}
{Department of Physics, Texas A\&M University \\
College Station, Texas 77843-4242, USA$^{1}$  \\
and \\
Laboratorium f\"{u}r Festk\"{o}rper Physik der ETHZ, ETH-H\"{o}nggerberg, 
8093, Zurich$^{2}$\\
and \\
Landau Institute for Theoretical Physics, Moscow, Russia$^{3}$
}
\end{instit}

\begin{abstract}
In 2D XY ferromagnet the dipole force induces a strong 
interaction
between spin-waves in the long-wavelength limit. The major 
effect of
this interaction is the transformation of a propagating spin-wave 
into a diffusion mode. We study the anomalous dynamics of 
such diffusion modes. We find that the Janssen-De Dominics 
functional, which governs this dynamics, approaches the
non-Gaussian fixed-point. A spin-wave propagates by an 
anomalous anisotropic diffusion with the dispersion relation: 
$i\omega{\sim}k_{y}^{\Delta_y}$ and $i\omega{\sim}k_{x}^{\Delta_x}$, 
where ${\Delta_y}=47/27$ and ${\Delta_x}=47/36$. The low-frequency 
response to the external magnetic field is found. 
\end{abstract}
\pagebreak

\section{Introduction}
\par
The notion of the ground state and independent elementary excitations lies 
in the background of the condensed matter theory. The excitations, 
such as electrons, holes, phonons, excitons, spin-waves etc. 
normally have a propagating, wave-like nature. In a homogeneous
medium they are characterized by their momentum or quasi-momentum
$\vec{\bf p}$ and their energy $\omega$. Each kind of excitations
has its specific dispersion relation or spectrum 
$\omega=\epsilon({\bf p})$. A wide scope of physical problems can be 
solved assuming the excitation to be independent, or considering their
interaction as a weak perturbation \cite{AGD}. However, in recent 
years more and more problems have occurred
to go beyond the simple picture of non-interacting or weakly interacting
excitations. In his pioneering work A.B.Migdal \cite{migdal} 
has indicated that electron-phonon interaction is not
weak in a narrow range of energy, leading to a strong renormalization
of the Fermi-velocity and even to an instability of the Fermi-surface.
\par
Recently a growing number of physical systems revealed
excitations which bare spectrum, obtained from linearized 
equations of motion, is strongly distorted by interaction
with vacuum and thermal fluctuations. A few
recent examples are: the so-called marginal Fermi-liquid, in which
electrons interacts via transverse magnetic field fluctuations 
\cite{correlectron}, and $\nu = 1/2$ state of the Fractional Quantum 
Hall effect in which initial electrons transform into quite 
different fermions \cite{stern}.
\par     
On the other hand, many years ago, the mode-mode interaction 
has been recognized as a necessary element of the critical dynamics 
\cite{kawasaki,kaswi,haho}. Particles,
heat and spin diffusion must be considered as hydrodynamic 
modes in the long-wavelength limit, as well as 
propagating waves, such as sounds, spin
waves etc. Their interaction has been proven to be substantial
not only in the critical region, but also for the hydrodynamics
of liquid crystals \cite{zeyher,kats} and for the CDW phason modes 
interacting 
with impurities \cite{gruner}. The excitation spectrum of these 
systems is reconstructed by effects of strong interaction.
   
Here we present a new solvable and experimentally feasible situation
where the strong interaction between spin-waves leads to the replacement
of the propagating spin-wave by a diffusion mode and to the appearance
of a new soft-mode in a certain range of momentum. This is the 
two-dimensional XY ferromagnet with dipolar interaction between 
spins.

The spin-diffusion mode appears naturally in the paramagnetic 
phase and in the vicinity of the Curie point \cite{Schwabl}. 
We consider, on the other hand, a low temperature ordered
phase, where no diffusion is expected but rather a propagating 
and weakly dissipating spin-wave mode. 

In a 2D XY ferromagnet at low temperatures the dipolar
interaction is relevant in the long wavelength limit,
even despite of the low density of spin-waves. It was
shown by one of the authors \cite{kashuba} that dipolar
force induces an anomalous anisotropic scaling of
spin-spin correlations in the ordered phase.
In this article we find an analogous dynamical scaling.

This paper is organized as follows. In the next section we
define the model and describe the spin-wave spectrum.
In section III we discuss the dynamics of XY-magnet 
with the dipole interaction and 
formulate the perturbation expansion for the model, using 
the Janssen-De Dominics \cite{martin} technique. 
Section IV is devoted to the solution of the Dyson
equation. There we find the self-induced dissipation of the spin-waves.
In section V the renormalization of the diffusion mode
is considered and the anomalous anisotropic exponents are found.
In section VI the dynamical susceptibility is found.
In conclusion we discuss prospects of the experimental observation of the 
anomalous modes.
In the Appendix A we use the Ward-Takahashi identities to prove that
the vertex corrections are small. Details of the Dyson equation solution 
can be found in the Appendixes B and C.
\par
A brief report on main results of this article has been published earlier
\cite{KAP}.

\section{The Hamiltonian and the spin-wave spectrum.}

\par
Spin-waves are fundamental excitations of the exchange magnet with 
spontaneously broken continuous symmetry. According to the so-called 
Adler principle \cite{kag-chub}, the interaction between 
spin waves vanishes in the long-wavelength limit. At low
temperatures the equilibrium density of spin waves is relatively small,
and a long wavelength non-equilibrium spin wave, excited by an 
external source, decays into other spin-waves 
or scatters on an equilibrium spin wave slowly. Thus, the dynamical
properties of the exchange magnet are determined by the well-defined
spin-wave mode. It also means that the imaginary part of the poles of 
the dynamical response function becomes much smaller than their 
real part as the wavelength grows to infinity.

The 2D exchange magnet with the easy-plane anisotropy is described 
by a following classical Hamiltonian:
\bee
H_{ExA}[{\bf S}]={\int}{\di}^{2}x\left[\frac{J}{2}
({\bf \nabla S})^{2}
+\frac{\lambda}{2}S_{z}^{2}\right]
-g_{G}{\mu}_{B}\int{\di}^{2}x{\bf S}{\bf H},
\ene
where $J$ is the exchange coupling constant, $\lambda$ is the 
strength of the easy-plane anisotropy $(\lambda >0)$ \cite{constants}, 
$g$ is the 
dipole interaction coupling constant, ${\bf H}$ is the
external magnetic field, $g_G$ is gyromagnetic ratio, 
${\mu}_B$ is the Bohr magneton and the field ${\bf S}({\bf x})$ 
represents the local spin of the magnet and can be normalized 
by a constraint 
\bee
{\bf S}^{2}({\bf x})=1.
\label{constraint}
\ene

The magnetic dipolar energy ${ H}_{dip}$ is represented by the sum:

\begin{equation} 
H_{dip}=\frac{g}{4\pi} \sum\limits_{{\bf x}_i \not= {\bf x}_j}
\frac{({\bf S}_i\cdot {\bf S}_j)-3({\bf S}_i\cdot {\hat \nu})
({\bf S}_j\cdot {\hat \nu})} 
{|{\bf x}_i-{\bf x}_j|^3}  
\end{equation}
where ${\hat \nu}$ is a unit vector pointing from ${\bf x}_i$ to ${\bf x}_j$,
${\bf S}_i={\bf S}({\bf x}_i)$, 
and $g={2\pi(g_G\mu_B Sa^{-2})^2}$ for the square lattice; for 
other lattices $g={2\pi(g_G\mu_B S\sigma)^2}$, where $\sigma$
is the inverse area of a plaquett of the lattice. 
The magnetic dipole energy can be separated into a short-range and 
a long-range part in the standard way \cite{KaP}. The short-range part 
renormalizes the single-ion spin anisotropy and favors in-plane 
spin orientation.  
The long-range part of the magnetic dipole energy is conveniently 
expressed in terms of the Fourier-transform ${\bf S}_{\bf k}$ of the
local magnetization field ${\bf S}({\bf x})$,  
\bee H_{dip}=
\frac{g}{2}\sum\limits_{\bf k}\frac{({\bf S_{k}k})
({\bf S}_{-{\bf k}}{\bf k})}{|{\bf k}|},\label{Hdip}
\ene
Thus, the total magnetic Hamiltonian is:
\bee H=H_{ExA}+H_{dip}. \label{Ham}
\ene  

\par
The dipole forces are crucial in the long-wavelength limit. Under the 
scaling transformation the short-range exchange 
interaction scales like $L^{d-2}$, whereas the dipole force scales
like $L^{2d-3}$, where $d=2,3$ is the spatial dimension of a magnet
and $L$ is the scale. The dipole energy is characterized by the 
dipole constant $g$ and the exchange energy is characterized by 
the exchange constant $J$. Normally in ferromagnets 
$ga^{d-1}$ is much smaller than $J$ ($a$ is the lattice constant).
However, as the scaling transformation shows, beyond the 
characteristic scale $L_d\sim (J/g)$ in 2D and $L_d\sim \sqrt{J/g}$ 
in 3D the dipole interaction dominates the energy of a spin-wave.  

\par
In the 2D $XY$ ferromagnet the dipole force stabilizes the 
long-range order \cite{maleev-pf}.
According to the Landau-Peierls-Hohenberg-Mermin-Wagner 
theorem (see e.g. \cite{HoM} ) in the absence of a long-range 
interaction 
the 2D magnet with broken continuous symmetry exhibits     
the algebraic decay of the spin correlations 
instead of the long-range order, and the infinite 
susceptibility at zero magnetic field. Thus, the dipole force 
plays a special role in 2D XY ferromagnet.  
In addition, the dipolar force is crucial for the spin statics
and dynamics in the ordered phase. In contrast to 
the 3D ferromagnet, where the spin waves are almost free, in 
2D the interaction between spin waves induced by dipole force
is dominant in the long-wavelength limit. 
\par
The Hamiltonian (\ref{Ham}) has two different scales: the anisotropy scale
$L_{A}=\sqrt{J/{\lambda}}$ and the dipole length $L_{D}=J/g$.
We assume that the anisotropy $\lambda$ is large compared to the 
dipolar energy ($L_A<L_D$). We direct the $y$-axis
along the net magnetization of the magnet 
and the $z$-axis perpendicular to the plane. 
The unit vector field ${\bf S}$ can be
represented by two scalar fields ${\phi}({\bf x},t)$ and 
${\cal \pi}({\bf x},t)$

\bee
{\bf S}=\left(-\sqrt{1-{\cal \pi}^2}\sin{\phi};
\sqrt{1-{\cal \pi}^{2}}\cos{\phi};
{\cal \pi}\right),                                             \label{S}
\ene
where both ${\cal \pi}$ and $\phi$ are small due to the fact that the
dipole force stabilizes the long-range order.

With the precision to the fourth power of $\phi$ and $\pi$ the 
Hamiltonian (\ref{Ham}) is:

\bea
H[\phi]=\frac{1}{2}{\int}_{\!\!\!{\bf k}}{\int}_{\!\!\! \omega}
\Biggl(
(J{\bf k}^{2}+h){\phi}_{{\bf k},\omega}{\phi}_{-{\bf k},-\omega}+
{\lambda}{\cal \pi}_{{\bf k},{\omega}}
{\pi}_{-{\bf k},-{\omega}}                  \nonumber       \\
+g\frac{\left(k_{x}(\phi_{{\bf k},\omega}-\phi^3_{{\bf k},\omega}/6)
+k_{y}\left[\frac{{\phi}^2}{2}\right]_{{\bf k},\omega}\right)
\left(k_{x}(\phi_{-{\bf k},-\omega}-\phi^3_{{\bf -k},-\omega}/6)+
k_{y}\left[\frac{{\phi}^2}{2}\right]_{-{\bf k},-\omega}\right)}
{|{\bf k}|}
\Biggr).                                              \label{Hamk}
\eea
Here 
${\phi}_{{\bf k},{\omega}}$ and ${\cal \pi}_{{\bf k},{\omega}}$
are the Fourier-transforms of the fields ${\phi}({\bf x},t)$ and 
${\cal \pi}({\bf x},t)$ respectively. We expanded the in-plane
magnetization components $cos\phi$ and $\sin\phi$ up to the 
fourth power in small spin fluctuations $\phi$.
We take the uniform magnetic field ${\bf H}$ to be directed
along the $y$-axis and $h=g_{G}{\mu}_{B}SH$.
The Fourier-transformed quantities are defined by

\bee
{\phi}({\bf x},t)={\int}_{\!\!\! \bf k}{\int}_{\!\!\! \omega}
{\phi}_{{\bf k},{\omega}}
e^{i({\bf k}{\bf x}-{\omega}t)},                             \label{FTrans}
\ene
where an abbreviated notation 
$$
{\int}_{\!\! \bf k}{\int}_{\!\! \omega}{\equiv}
{\int}{\!}{\int}\frac{{\di}^{2}k}{(2{\pi})^2}\frac{\di \omega}{2{\pi}}
$$
is used and 
$\left[\frac{{\phi}^2}{2}\right]_{{\bf k},{\omega}}$ denotes the 
Fourier-transformation of $\frac{{\phi}^2({\bf x},t)}{2}$.

\par
The interaction between the spin waves is described by the non-quadratic
part of the Hamiltonian (\ref{Hamk}):

\bea
H_{\mbox{int}}={\int}_{\!\! {\bf k}_1 ,{\bf k}_2 ,{\bf k}_3}
{\int}_{\!\! \omega_1,\omega_2,\omega_3}
f({\bf k}_1 ,{\bf k}_2 ,{\bf k}_3)
\phi_{{\bf k}_1 ,\omega_1} \phi_{{\bf k}_2 ,\omega_2} 
\phi_{{\bf k}_3 ,\omega_3}                               \nonumber  \\
\times\delta({\bf k}_1 +{\bf k}_2 +{\bf k}_3)
\delta(\omega_1 +\omega_2 +\omega_3)                     \nonumber  \\  
+{\int}_{\!\! {\bf k}_1 ,{\bf k}_2 ,{\bf k}_3 ,{\bf k}_4}
{\int}_{\!\! \omega_1,\omega_2,\omega_3,\omega_4} 
g({\bf k}_1 ,{\bf k}_2 ,{\bf k}_3 ,{\bf k}_4)
\phi_{{\bf k}_1 ,\omega_1} \phi_{{\bf k}_2 ,\omega_2} 
\phi_{{\bf k}_3 ,\omega_3} \phi_{{\bf k}_4 ,\omega_4}    \nonumber   \\
{\times}\delta({\bf k}_1 +{\bf k}_2 +{\bf k}_3+{\bf k}_4)
\delta(\omega_1 +\omega_2 +\omega_3+\omega_4),           \label{interaction}
\eea
where the three-leg bare vertex $f({\bf k}_1 ,{\bf k}_2 ,{\bf k}_3)$ is:

\bee
f({\bf k}_1 ,{\bf k}_2 ,{\bf k}_3)=\frac{g}{3}
\sum\limits_{i=1}^{3}\frac{k_{ix}k_{iy}}{|{\bf k}_i|}    \label{threeleg}
\ene
and the four-leg vertex is defined as:

\bee
g({\bf k}_1 ,{\bf k}_2 ,{\bf k}_3 ,{\bf k}_4)=
\frac{g}{24}\sum\limits_{i>j=1}^{4}
\frac{(k_{iy}+k_{jy})^2-(k_{ix}+k_{jx})^2}{|{\bf k}_i +{\bf k}_j|}.
       \label{fourleg}                    \ene

\par
The vertices (\ref{threeleg}) and (\ref{fourleg}) decrease as the 
momentum ${\bf k}$ goes to zero. Besides of that, these 
vertices are singular at $|{\bf k}| \rightarrow 0$. Nevertheless,
the interaction between spin waves asymptotically vanishes in the
long wavelength limit for a 3D magnet \cite{kag-chub}. No renormalization
of the bare correlator

\bee
K({\bf k})=\langle \delta S_x({\bf k}) 
\delta S_x(-{\bf k}) \rangle = 
\langle \phi_{\bf k} \phi_{-{\bf k}} \rangle =
\frac{T}{J{\bf k}^2+g\frac{k_x^2}{|{\bf k}|}}         \label{barecorr}
\ene
appears in this limit (we put $h=0$).

\par
In 2D the situation changes drastically: the interaction grows
with the wavelength, resulting in strong renormalization of critical
exponents. To show the difference we calculate the upper marginal 
dimension of the Hamiltonian (\ref{Hamk}). Let us consider an arbitrary
diagram from the perturbation expansion of some correlator.
In order to add an internal line to such a diagram, we need to add
three bare correlators $K$ given by  eq. (\ref{barecorr}), 
two vertices $f$ and one integration over $\di k_x \di^{(D-1)} k_y$. 
From eq. (\ref{barecorr})
we see that for small momenta, $k_x \sim k_y^{3/2}$ and  
$K \sim k_y^{-2}$. Equation (\ref{threeleg}) gives 
$f \sim k_x \sim k_y^{3/2}$. Hence, if we require that the one-line insertion
be dimensionless, we obtain: $3\cdot (-2)+2\cdot 3/2+3/2+(D-1)=0$ 
or $D=5/2$. It means that the theory is renormalizable in $D \leq 5/2$.
It turns out that in the static 2D case the critical exponents can be 
found exactly.

\par
Following the work \cite{kashuba}, we rescale the field 
$\phi^2 \rightarrow \phi^2/\sqrt{Jg}$ and rewrite the Hamiltonian
(\ref{Hamk}) in a slightly different form:

\bea
H=\int_{\! \bf k}\left(ak^2_y \phi_{\bf k} \phi_{-{\bf k}}+
a^{-1}\frac{k_x^2}{|k_y|}\phi_{\bf k}\phi_{-{\bf k}}+
2w\frac{k_x k_y}{|k_y|}\phi_{\bf k}
\left[\frac{\phi^2}{2}\right]_{-{\bf k}}+
w^2 a|k_y|\left[\frac{\phi^2}{2}\right]_{\bf k}
\left[\frac{\phi^2}{2}\right]_{-{\bf k}}
\right),                                            \label{KashHam}
\eea 
where $a=\sqrt{J/g}$, $w^2=\sqrt{g/J^3}$ and we have taken
into account that for small ${\bf k}$ one can substitute 
$|{\bf k}|\rightarrow k_y$ (see (\ref{barecorr})).
Requiring that the Hamiltonian (\ref{KashHam}) does not change
under scale transformation $k_y \rightarrow lk_y$, 
$k_x \rightarrow l^{\Delta_x^0}k_x$, 
$\phi_{\bf k} \rightarrow l^{\Delta_{\phi}^0}\phi_{\bf k}$,
$a \rightarrow l^{\Delta_a^0}a$, $w \rightarrow l^{\Delta^0}w$,
we find the bare exponents $\Delta_x^0=3/2$, $\Delta_a^0=0$,
$\Delta_{\phi}^0=-9/4$, $\Delta_w^0=1/4$ (where we have taken into account
that $\left[\frac{\phi^2}{2}\right]$ has one more integration). 
Now, according to the
standard procedure \cite{zinn}, we introduce 
the renormalized field and charges: 
$\phi^2=Z_{\phi}\phi_R^2$, $a=Z_a a_R$, $w=Z_w \mu^{1/4}\tilde{w}_R$,
where $\mu$ is the scale at which the dimensionless 
$\tilde{w}_R$ is observed. The second term in the Hamiltonian 
(\ref{KashHam}) is non-analytic at small momenta. It means that 
there can be no corrections to this term in the regular perturbation 
expansion. Furthermore, from Appendix A one can see, that in this case, 
due to symmetry, there will be no correction to the third term either.
Hence, we can establish the relations between the renormalization 
coefficients:  

\bea
Z_a^{-1}Z_{\phi}=1                     \label{ZaZp}    \\
Z_{\phi}^{3/2}Z_w \mu^{1/4}=1.         \label{ZpZw}
\eea
It can be proven that the renormalization group flow has a 
non-zero fixed point. It means that $\tilde{w}_R=$const and hence
$Z_w \sim (\Lambda/\mu)^{1/4}$, where $\Lambda$ is the scale at which
the coupling constant equals $w$. Introducing the critical exponents
$Z_{\phi}\sim \Lambda^{2\Delta_{\phi}}$, $Z_a\sim \Lambda^{\Delta_a}$,
we find from (\ref{ZaZp}) and (\ref{ZpZw}) that $\Delta_{\phi}=-1/12$
and $\Delta_a=-1/6$. Demanding that the first and the second terms in the
Hamiltonian (\ref{KashHam}) have the same dimensions, we find
$\Delta_x=4/3$. Finally, from the Callan-Symanzik equation \cite{zinn},
the long range limit of the two point correlation function can be found:

$$
K({\bf x})=(x^2+|y|^{8/3})^{-1/4}f(x/|y|^{4/3}),
$$
where $f(x)$ is an arbitrary finite function. 
The dynamical properties of the magnet, however, are still to be found. 

\section{The Dynamics of XY Magnet}
\par
At $T=0$ the classical 
magnet with the Hamiltonian~(\ref{Ham}), obeying the 
constraint (\ref{constraint}), follows the Landau-Lifshitz equations 
(see e.g. \cite{LL}):

\bee
S{\hbar}\frac{\partial {\bf S}({\bf x},t)}{\partial t}=
{\bf S}({\bf x},t){\times}
\frac{\delta H}{\delta {\bf S}({\bf x},t)}\, ,     \label{LL}
\ene
where $S$ is the absolute value of a spin localized on a magnetic ion.
In terms of the canonically conjugated fields $\pi$ and $\phi$,
eq.~(\ref{LL}) can 
be rewritten in the Hamiltonian form:

\bea
{\hbar}S{\partial}_{t}{\cal \pi}({\bf x},t)=
\frac{{\delta}H}{{\delta}{\phi}({\bf x},t)}\, ,          \label{Eqnpi}  \\  
-{\hbar}S{\partial}_{t}{\phi}({\bf x},t)=
\frac{{\delta}H}{{\delta}{\cal \pi}({\bf x},t)}\approx
{\lambda}{\cal \pi}({\bf x},t).               \label{eqnphi}
\eea

\par
In the harmonic approximation equations (\ref{Eqnpi},\ref{eqnphi}) imply 
the dispersion relation for the spin-wave mode~\cite{constants} in a XY
magnet:

\bee
{\epsilon}^{2}({\bf k})=
{\lambda}\left(J{\bf k}^{2}+g\frac{k_{x}^{2}}{|{\bf k}|}\right)=
c^{2}\left({\bf k}^{2}+p_{0}\frac{k_{x}^{2}}{|{\bf k}|}\right),
\label{disp}
\ene
where $c=\sqrt{\lambda J}$ is the spin-wave velocity and $p_{0}=g/J$.
The out-of-plane anisotropy $\lambda$ affects the dynamics in the 
long-wavelength limit $k \ll p_{\lambda}=\sqrt{\lambda/J}$ 
considered in this article. Thus, $p_{\lambda}$ is the 
upper cut-off momentum in our theory. A spin wave with $k \gg p_0$
has the phonon-like isotropic spectrum $\epsilon=ck$. The range
$p_0 \ll k \ll p_{\lambda}$ will be called acoustic shell 
(${\cal A}$-shell). At lower momenta $p \ll p_0$ the spin-wave 
spectrum is dominated by the dipolar interaction: 
$\epsilon({\bf k})\approx c\sqrt{p_0 k}\sin \theta$, where $\theta$ 
is the angle between the direction of the spontaneous magnetization
and the wave-vector. This range of momenta will be called dipolar
shell (${\cal D}$-shell). The effect of the presence of the dipolar
force in the 2D XY-magnet is not limited to the change of free 
spin-wave spectrum. As we have mentioned earlier, it leads to  
strong spin-wave interaction and to a crucial transformation of the 
spin propagation.

\par                   
Without dipolar forces the dynamics of the 2D XY ferromagnet is 
well described by non-interacting spin waves. The
high level of fluctuations leads to a strong temperature dependence
of the dynamic spin correlators, which have algebraic character,
just as static ones \cite{PU,Ber}.
The renormalized two-point spin correlation function features
the pole with the temperature dependent power exponent.
The dipole force suppresses such strong spin fluctuations, 
but not entirely.

\par 
In addition, the dipolar interaction induces decay processes.
As a result finite spin-wave life-time $\Gamma({\bf k})$ or
the width of the level $b({\bf k})=\Gamma^{-1}({\bf k})$ appears.
In 3D at low temperature and at small momentum $|{\bf k}|$, the width 
$b({\bf k})$ is much smaller than $\omega({\bf k})$. In 2D, however, 
the interaction is essential and must be considered seriously.  

\par
To take into account the dissipation induced by thermal fluctuations 
at a temperature $T$, we introduce a phenomenological dissipation
functional~\cite{Rfunction}:

\bee
R[\phi]={\int}{\di}t{\di}^{2}x{\di}^{2}x'R({\bf x}-{\bf x}')
\dot{\phi}({\bf x})\dot{\phi}({\bf x}').                   \label{relax}
\ene 
Eliminating ${\pi}$ from equations (\ref{Eqnpi},\ref{eqnphi}) 
and adding a proper dissipation term, one obtains a following equation 
for $\phi({\bf x},t)$:

\bee
-\frac{1}{\lambda}{\partial}^{2}_{t}{\phi}({\bf x},t)=
\left[ \frac{1}{\lambda} \epsilon^{2}({\bf k})
{\phi}({\omega},{\bf k})\right]_{{\bf x},t}+
\frac{{\delta}H_{\mbox{int}}}{{\delta}{\phi}({\bf x},t)}+
\frac{\delta R}{\delta \dot{\phi}({\bf x},t)}+
{\eta}({\bf x},t)-h({\bf x},t)\, ,       \label{eqn}
\ene
where $h({\bf x},t)$ is the external magnetic field, and the 
interaction part of the Hamiltonian is determined by equation
(\ref{interaction}).
We have introduced the random noise 
$\eta({\bf x},t)$ in equation (\ref{eqn}). The noise, 
in effect, generates dissipation. As usual, the random noise is assumed to
obey the Gaussian statistics. Its correlation function is determined by the
fluctuation-dissipation theorem \cite{LL5}:

\bee
{\langle}{\eta}_{\bf k}{\eta}_{-{\bf k}}{\rangle}=2TR({\bf k}) \label{noise}
\ene
Here $R({\bf k})$ is the Fourier-transform of the
function $R({\bf x}-{\bf x}')$.
The dissipation in the exchange ferromagnet vanishes in the long-wavelength
limit \cite{kag-chub}: $R({\bf k})=b{\bf k}^2$.
In 2D XY dipole magnet the dissipation 
does not vanish in the long-wavelength limit. We notify:

\bee
R({\bf k})=\frac{1}{\Gamma ({\bf k})}               \label{second}
\ene

\par
We emphasize that the finite life-time $\Gamma$ is determined 
self-consistently by the processes of the decay and scattering of 
spin waves. We neglect the spin-wave-electron and spin-wave-sound
interactions. The first interaction is not weak, but the Fermi-velocity
is much higher than the spin-wave velocity and the spin-electron 
interaction is not effective
for long wavelength. Even the sound velocity can be much larger than
the spin-wave velocity, since the latter is proportional to small
$\sqrt{\lambda}$. 

\par
Our aim then is to calculate the linear response function $G({\bf x},t)$
to a weak external field $h({\bf x},t)$ (see equation (\ref{eqn}))
averaged over the thermal fluctuations $\eta ({\bf x},t)$.
We apply the Janssen-De Dominicis functional method \cite{martin} to
reformulate stochastic  equation~(\ref{eqn}) in terms of the path 
integral. The
probability distribution for the noise ${\eta}({\bf x},t)$ is: 

\bee
W[{\eta}]{\sim}{\exp}{\left[-\frac{1}{4T}{\int}{\di}^{2}x{\di}^{2}x'
{\int}{\di}t{\eta}({\bf x},t)R^{-1}({\bf x}-{\bf x}')
{\eta}({\bf x}',t)\right]}.                                  \label{noisedis}
\ene

\par
Following the standard dynamic field theory \cite{zinn,kats}, upon 
averaging over the noise distribution and introducing auxiliary 
response field
${\phe}({\bf x},{t})$, one can reduce the solution of the stochastic 
differential equation~(\ref{eqn}) to the calculation of the dynamical 
partition function:
\bee
Z[j,\jh]={\int}{\cal D}[\phi]{\cal D}[i\phe]{\exp}{\left(
{\cal J}[{\phi},{\phe}]+{\int}{\di}^{2}x{\int}{\di}t
{\left[{\jh}{\phe}+j{\phi}\right]}\right)}                   \label{Zfunc}
\ene
and its derivatives over the currents $j$ and $\jh$. 
Here ${\cal J}({\phi},{\phe})$ is
the Janssen-De Dominics functional (JDF).

\bea
{\cal J}[\phi,\phe]={\int}{\di}^{2}x{\int}{\di}^{2}x'{\int}{\di}t
{\phe}({\bf x},t)TR({\bf x}-{\bf x'}){\phe}({\bf x}',t)   \cr
-{\phe}({\bf x}',t){\left[\frac{1}{\lambda}{\partial}^{2}_{t}{\phi}({\bf x},t)
{\delta}({\bf x}'-{\bf x})+\frac{\delta H}{\delta {\phi}({\bf x},t)}
{\delta}({\bf x}'-{\bf x})+R({\bf x}'-{\bf x}){\partial}_{t}
{\phi}({\bf x},t)\right]}.                                       \label{JDfunc}
\eea
By differentiation of the JDF over $j$ and $\hat{j}$ one can obtain any
correlation function. 
\par
We define the inverse bare propagator ${\hat {\bf G}}_0^{-1}$

\begin{eqnarray}
{\hat {\bf G}}_{0}^{-1}=
\left( 
\begin{array}{cc}
-2TR({\bf k}) & \frac{\omega^{2}}{\lambda}
-\frac{\epsilon^{2}({\bf k})}{\lambda}-
i{\omega}R({\bf k}) \\
\frac{{\omega}^{2}}{\lambda}
-\frac{\epsilon^{2}({\bf k})}{\lambda}+ i{\omega}
R({\bf k}) & 0
\end{array} 
\right)\, ,                                                  \label{K0}
\end{eqnarray}
so that the harmonic ("Gaussian") part of ${\cal J}$ can be written in the
form:

\bea
{\cal J}_{0}[\phi,\phe]=-\frac{1}{2}{\int}_{\!\!\! \bf k}{\int}
_{\!\!\! \omega}
\left(
\begin{array}{cc}
{\hat \phi}({\omega},{\bf k}), & {\phi}({\omega},{\bf k})
\end{array}
\right)
{\hat {\bf G}}_{0}^{-1}({\omega},{\bf k})
\left(
\begin{array}{c}
{\phe}(-{\omega},-{\bf k}) \\
{\phi}(-{\omega},-{\bf k})
\end{array}
\right) .                                                  \label{J0}
\eea

\par
From eq. (\ref{K0}) one finds the bare propagator:

\bee
{\hat {\bf G}}_{0}=
\left(
\begin{array}{cc}
0 & G_{0}^{\ast}({\omega},{\bf k})\\
G_{0}({\omega},{\bf k}) & 
D_{0}({\omega},{\bf k})
\end{array}
\right)\, ,                                                   \label{GMatrix}
\ene
where we define the bare dynamical response and spin-spin correlation 
function as follows: 

\bea
G_{0}({\omega},{\bf k})=
\frac{\lambda}{{\omega}^{2}-\epsilon^{2}({\bf k})
-i{\omega}{\lambda}R({\bf k})}\, ,                          \label{GFunc}  \\
D_{0}({\omega},{\bf k})=
\frac{2T \lambda R({\bf k})} {
\left[ \omega^{2}-\epsilon^{2}({\bf k}) \right]^{2}
+{\omega}^{2}{\lambda^2}R^{2}({\bf k}).
}                                                          \label{DFunc}
\eea

They obey the standard fluctuation-dissipation relation:
$D_0=\frac{2T}{\omega}{\cal I}mG_0$. The same relation is correct
for the total dynamic correlation $D(\omega,{\bf k})$ and the total
linear response function $G(\omega,{\bf k})$: 
$D=\frac{2T}{\omega}{\cal I}mG$. 

\par
The anharmonic (interaction) part of ${\cal J}$ is:

\bea
&{\cal J}_{\mbox{int}}&=
\sum\limits_{{\omega},{\bf k}}{\phe}_{\omega,{\bf k}}
\frac{\delta H_{\mbox{int}}}{\delta \phi_{\omega,{\bf k}}} 
=\sum\limits_{{\omega},{\bf k}}{\phe}_{\omega,{\bf k}}
\Biggl(3\sum\limits_{{\omega}_2,\omega_3}
\sum\limits_{{\bf k}_2,{\bf k}_3}
f({\bf k},{\bf k}_2 ,{\bf k}_3)
\phi_{{\bf k}_2 ,\omega_2} 
\phi_{{\bf k}_3 ,\omega_3}                               \nonumber  \\
&\times& \delta({\bf k} +{\bf k}_2 +{\bf k}_3)
\delta(\omega +\omega_2 +\omega_3)                     
+4\!\!\!\sum\limits_{{\omega}_2,\omega_3,\omega_4}
\sum\limits_{{\bf k}_2,{\bf k}_3,{\bf k}_4}
g({\bf k} ,{\bf k}_2 ,{\bf k}_3 ,{\bf k}_4)
\phi_{{\bf k}_2 ,\omega_2} 
\phi_{{\bf k}_3 ,\omega_3} \phi_{{\bf k}_4 ,\omega_4}    \nonumber   \\
&{\times}&\delta({\bf k} +{\bf k}_2 +{\bf k}_3+{\bf k}_4)
\delta(\omega +\omega_2 +\omega_3+\omega_4)\Biggr),     \label{Jinteraction}
\eea

\par
We define in a common way the self-energy operator
${\hat {\bf \Sigma}}(\omega,{\bf k})$ by the relation:

\bee
{\hat{\bf G}}^{-1}({\omega},{\bf p})=
{\hat{\bf G}}_0^{-1}({\omega},{\bf p})
-{\hat {\bf {\Sigma}}}({\omega},{\bf p}),                 \label{Ghat}
\ene

where:

\bee
{\hat {\bf G}}=
\left(
\begin{array}{cc}
0 & G^{\ast}({\omega},{\bf k})\\
G({\omega},{\bf k}) & 
D({\omega},{\bf k})
\end{array}
\right)                                                    \label{GMatrixTot}
\ene
and $G(\omega,{\bf k})$ and $D(\omega,{\bf k})$ are the complete
response function and correlator respectively.

\par
The self-energy ${\hat {\bf \Sigma}}(\omega,{\bf k})$ 
satisfies Dyson equation:
\bee
{\hat {\bf {\Sigma}}}({\omega},{\bf k})=g{\int}_{\!\!\! \Omega}
{\int}_{\!\!\! {\bf p}}{\hat{\bf G}}({\Omega},{\bf p})
{\hat \Lambda} ({\bf p, k};\omega ,\Omega) 
{\hat{\bf G}}({\omega}-{\Omega},{\bf k}-{\bf p}),  \label{sigma}
\ene
where ${\hat \Lambda} ({\bf p, k};\omega ,\Omega)$ is the full vertex. 
The bare vertex is given by equation:

\bee
{\hat \Lambda}_0 ({\bf p},{\bf k};\omega ,\Omega) = 
\frac{2TR({\bf p})}{\lambda}f^2 ({\bf p},{\bf k}) 
\left(
\begin{array}{cc}
-D_0(\Omega,{\bf p}) & G^{\ast}_0({\Omega},{\bf p})\\
G_0({\Omega},{\bf p}) & 0
\end{array}
\right).                                                    \label{LMatrix0}
\ene
We have denoted $f({\bf p},{\bf k})\equiv 
f({\bf p},{\bf k},{\bf p}-{\bf k})$.

Further we consider a limit $R({\bf k}){\rightarrow}+0$. 
According to the FDT, the matrix 
${\hat {\bf \Sigma}}(\omega,{\bf p})$ must have a form:

\bee
{\hat {\bf \Sigma}}(\omega,{\bf p})=
\left(
\begin{array}{cc}
\frac{2T}{\lambda \omega} {\cal I}m{\Sigma}(\omega,{\bf p}) & 
\frac{1}{\lambda} {\Sigma}^{\ast}(\omega,{\bf p})   \\
\frac{1}{\lambda} {\Sigma}(\omega,{\bf p}) & 0
\end{array}
\right),                                            \label{SigmaMatrix}
\ene
where the self-energy function $\Sigma (\omega,{\bf p})$ is associated
with the complete response function $G(\omega,{\bf p})$ by the same 
relationship:

\bee
G^{-1}(\omega,{\bf p})=G_0^{-1}(\omega,{\bf p})
-\frac1{\lambda}\Sigma(\omega,{\bf p}).                  \nonumber
\ene

\par
In the Appendix A we show that the full vertex $\Lambda$ 
can be approximated with a good accuracy by its bare 
value $f(\omega,{\bf p})$ in the low-frequency range.
In this respect our theory reminds the Migdal theory of 
the interacting electron-phonon system \cite{migdal}. 
In the Migdal theory the simplification is due to a 
narrow scale of the energy shell in which the interaction 
proceeds. In our theory we assume that the frequency of 
spin fluctuations is small instead.

We notify the real and the imaginary part
of the self-energy term as: $\Sigma =a^2(\omega,{\bf p})
-i\omega b(\omega,{\bf p})$. Thus, the Green function (\ref{Ghat})
reads:

\bee 
G^{-1}(\omega,{\bf p})=\omega^2-\epsilon^2
({\bf p})-a^2(\omega,{\bf p})+i\omega b(\omega,{\bf p}),  \nonumber
\ene
while the spin-spin correlation function is:

\bee 
D(\omega,{\bf p})=\frac{b(\omega,{\bf p})}{\left[
\omega^2-\epsilon^2({\bf p})-a^2(\omega,{\bf p})
\right]^2+\omega^2b^2 (\omega,{\bf p})}                 \nonumber
\ene
(we have slightly changed the definitions of $G$ and $D$ and 
referred the factor $\sqrt{2T\lambda^3}$ to the vertex).

\par
We employ the reduced temperature $t=T/4\pi J$
and the ratio $g/\sqrt{J\lambda}=p_0/\sqrt{\lambda/J}$ 
as small parameters. The latter means that the 
${\cal A}$-shell is much larger than the ${\cal D}$-shell.
We also use the notation $L=\log(\sqrt{J\lambda}/g)$.

\par
The main contribution to the
self-energy is given by the one-loop diagrams shown 
in Fig.1a, 1b. 
Our theory is valid only if the temperature is small: 
\bee t\log(\sqrt{J\lambda}/g)=tL\ll 1. \label{tcon} \ene
Under this condition the two-loop corrections are small, and
the diagram Fig.1b 
contributes to a negligible change
of the spectrum (\ref{disp}) \cite{zinn}. Such neglecting 
of the two-loop diagrams (vertex correction) was a major
assumption in the so-called mode-coupling methods 
\cite{Schwabl}. This approximation serves well in the theory
of the 3D critical dynamics with the dipole force being included. 
Later we prove this assumption for 2D.
Thus, the Dyson equation for our problem is as follows:

\bee
\Sigma(\Omega,{\bf q})=18\lambda^3
T\int_{\! {\bf p}}\int_{\! \omega}
f^2({\bf p},{\bf q}) 
D(\omega,{\bf p})G(\omega+\Omega,{\bf q}-{\bf p}),           \label{rea}
\ene

\par
The functions
$b(\omega,{\bf p})$ and $a(\omega,{\bf p})$ are even
in both arguments \cite{AGD}.
The imaginary part of the self-energy 
is odd in $\omega$:
${\cal I}m\Sigma(\Omega,{\bf q})=-\Omega b(\Omega,{\bf q})$. Hence,
equation for the dissipation function reads:

\bee
b(\Omega,{\bf q})=9\lambda^3T
\int_{\!\! {\bf p}}\int_{\! \omega}
f^2({\bf p},{\bf q})
D(\omega,{\bf p})D(\omega+\Omega,{\bf q}-{\bf p}).             \label{ME}
\ene

\par
The integrand in (\ref{ME}) is positive. Thus, the main 
contribution to $b(\Omega,{\bf q})$ comes from the region, where 
poles of the two $D$-functions coincide. The function 
$D(\omega,{\bf p})$ has poles at 
$\omega\approx\pm\epsilon({\bf p})$ in the
${\cal A}$-shell. Following the terminology of the field 
theory, we call the surface $\omega^2=\epsilon^2({\bf p})$ the 
mass-shell. The self-energy in the ${\cal A}$-shell is small as 
it is shown in the Appendix B and we neglect it. Because the
dissipation is small, the $D$-function can be represented as a 
sum of $\delta$-functions:
\bee 
D(\omega,{\bf p})\approx\sum_\pm
{\pi\over 2\epsilon^2({\bf p})}
\delta(\Delta\omega_\pm),                           \label{D1} 
\ene
where $\Delta\omega_\pm=\omega\pm\epsilon({\bf p})$
measures the deviation from the mass-shell.
After integrating $\omega$ out from eq. (\ref{ME})
with the $D$-functions from (\ref{D1}), we recover the
Fermi Golden Rule for the probability of the spin-wave
decay and scattering processes.

\par
Looking for the long wavelength quasi-excitations, we need 
the self-energy at very small momenta $q\ll p_0$, which we 
denote as $\Sigma_0$. We expect the quasi-excitations to be soft:
$\Omega\ll cq$. Here we restrict the quasi-excitation wave-vector 
${\bf q}$ to be directed almost along the magnetization: 
$|q_x|\ll q$ (arbitrarily directed ${\bf q}$ are considered in Appendix 
C). The essential contribution to the integral in equation (\ref{rea})
comes from the internal momentum $p$ being in the $\cal A$-shell
and the internal frequency $\omega=\epsilon({\bf p})$. 
Performing the integration over $\omega$ with the $D$-function 
from equation (\ref{D1}), we find (for derivation see Apendix B and
note that we neglect the real part of the self-energy in the 
${\cal A}$-shell, as it is justified in the Appendix B):
\begin{equation}
\Sigma_0={c^2p_0^2t\over 4\pi}\int
{c^4p^3\di p\over\epsilon^4({\bf p})}{\Omega\sin^2(2\psi)
d\psi\over \Omega-cq\cos\psi+ib_1},                     \label{B0}
\end{equation}
where $b_1$ is the dissipation function $b(\omega,{\bf p})$ 
of a spin-wave inside the ${\cal A}$-shell. Note, that
${\cal R}e\Sigma_0$ vanishes in the static limit
$\Omega =0$.

\par
If $cq\gg b_1$, we make the integral over $\psi$ in 
equation (\ref{B0}) to find:

\bee  
\Sigma_0(\chi)=c^2p_0^2tL
\cos^2\chi\exp({-2i\chi}),                             \label{B00} 
\ene
where $\chi$, defined by the equation $\cos\chi=\Omega/cq$,
measures the deviation from the mass shell.
More generally, we introduce a notation $r$ for the
ratio $\Omega/cq$ ($r=\cos \chi$ if $r\leq 1$). Then

\bee
\Sigma_0(r)=c^2 p_0^2 L t r^2(2r^2-1-2r\sqrt{r^2-1}).  \label{B00r}
\ene
Note that $\Sigma_0(r)\approx -c^2 p_0^2 t/4$ when 
$r\rightarrow\infty$ and $|\Sigma_0(r)|<c^2 p_0^2 t/4$ 
at any $r$. The self-energy $\Sigma_0(r)$ is real for $r>1$.

\par
If $q$ is so small that $cq\ll b_1$, eq. (\ref{B0}) 
implies the $q$-independent dissipation constant:

\bee 
b_0=c^2p_0^2 t L\int{\di\psi\over 4\pi}
{\sin^2(2\psi)\over b_1(\psi)}.                        \label{MR} 
\ene
In this calculation we have used the fact that the
dissipation of a spin-wave in the ${\cal A}$-shell $b_1$ 
depends only on the angle $\psi$ between the direction 
of magnetization and the spin-wave wave-vector ${\bf p}$ 
which we prove below.

\par
Now we need to calculate $b_1(\psi)$. An unusual feature of our
theory is that the dissipation process in the
${\cal A}$-shell is mediated by an off-mass-shell
virtual spin wave. Indeed, the dispersion relation (\ref{disp})
does not allow for decay or merging processes.
Alternatively, as we will show, the dissipation of a
spin-wave in the ${\cal A}$-shell, propagating along the
direction specified with the angle $\psi$ 
($\sin\psi=q_x/q$), is mediated by an internal virtual 
spin-wave in (9), with a momentum of $p\ll p_0$ and a 
frequency of $\omega <cp$, propagating along the direction 
very close to the $y$-axis $\vphi^2\ll 1$ 
(where $\sin\vphi=p_x/p$), 
to provide a finite attenuation 
of this state. The integration over $\omega$ with one of 
the $D$-functions in (\ref{ME}), taken in the form 
(\ref{D1}), leads to a following equation:

\bee 
b_1=-9c^4t\frac{f^2({\bf 0},{\bf q})}{8\pi J^2 q^2}
\int \di^2{\bf p}D(\epsilon({\bf p+q})-
\epsilon({\bf q}),{\bf p}).                         \label{sss} 
\ene
Since $\omega=\epsilon({\bf p+q})-\epsilon({\bf q})$,
we conclude that $\omega=cp\cos\Phi$, where
$\Phi=\psi-\vphi\approx\psi$ is the angle between the vectors
${\bf q}$ and ${\bf p}$. 
According to fluctuation-dissipation theorem: 
$D(\epsilon_{\bf p+q}-\epsilon_{\bf q},{\bf p})=
\frac{1}{\epsilon_{\bf p+q}-\epsilon_{\bf q}}{\cal I}m
G(\epsilon_{\bf p+q}-\epsilon_{\bf q},{\bf p})$.
Invoking the definition of the angle $\chi$ for virtual
spin-wave, we find that $\chi=\Phi\approx\psi$. Substituting
$f^2({\bf 0},{\bf q})=(1/9)\,q^2\sin^2(2\psi)$, 
$\epsilon^2({\bf p})=c^2p^2+c^2p_0p\sin^2\vphi
\approx c^2p^2+c^2p_0p\vphi^2$ and taking into account that
$\Sigma_0$ from (\ref{B00}) depends only on $\chi=\psi$, we
write: 

\bee 
b_1(\psi)={c^2p_0^2t\over 2\pi}{\cal I}m\int
\frac{\sin^2\psi\cos\psi\,\di p\di\vphi}{p^2\sin^2\psi+
p_0p\vphi^2+\Sigma_0(\psi)/c^2}.                     \label{B1}
\ene
Note that the most dangerous region of integration is the region of
small $p$, such that $\vphi^2\sim p/p_0\ll 1$ in (\ref{B1}).
In other words, the
dissipation of a short wavelength spin-wave, propagating
in the direction $\psi$, is determined by the scattering
on the long wavelength virtual spin-wave, with the momentum along 
${\hat y}$-direction, which lies
on a specific distance off the mass-shell: $\omega/cp=
\cos\psi$. The integration over $p$ in (\ref{B1})
is confined towards the crossover region: 
$p\sim p_c=p_0\sqrt{ t L}$.

\par
To find  the anisotropic dissipation of a 
spin-wave mode in the ${\cal A}$-shell, we 
plug $\Sigma_0(\psi)$ from equation (\ref{B00}) into 
(\ref{B1}). After a change of variables 
$(p,\varphi)\rightarrow(\rho,\vartheta)$, given by 
formulae $p=p_0\rho^2\cos\vartheta$ and 
$\varphi=\rho\sin\psi\sin\vartheta\cos^{-1/2}\vartheta$
($-\infty<\rho<\infty$ and $0<\vartheta<\pi/2$),
the integration becomes trivial and gives:

\bee 
b_1(\psi)=\beta_1 t^{3/4}cp_0{\sin^{3/2}(2\psi)
\sin(\psi/2)\over L^{1/4}\cos\psi},                    \label{B11}
\ene
where the direction of the spin-wave is limited to
the fundamental quadrant: $0<\psi<\pi/2$, and
$\beta_1=\Gamma^2(1/4)/4\sqrt{2\pi}\approx 1.31$.

\par
Let us return to the range of very low momenta 
$p\ll b_1/c$. Plugging (\ref{B11}) into (\ref{MR}), 
one finds:
$$ b_0=\beta_0 cp_0t^{1/4}L^{5/4},$$ 
where $\beta_0\approx 1.24$.
The condition $cp_{DM}\sim b_1$ defines the
crossover wave-vector: $p_{DM}\sim \beta_1 p_0t^{3/4}/L^{1/4}$,
between the self-energies (\ref{B00}) and (\ref{MR}).
The dissipation functions (\ref{B00},\ref{MR} and \ref{B11}) 
represent the self-consistent solution of the Dyson equation 
(\ref{rea}, \ref{ME}).

\par
Finally, we verify that the two-loop correction (see Fig.1c)
is negligible. There exists several diagrams with different 
arrangements of $G$ and $D$ functions. On each of the two short 
loops on the diagram Fig.1c there exists at least one $D$ function 
but there may be two of them. We consider only the most 'dangerous' 
diagram with each short loop having exactly one $D$-function (see
Fig.2). Note that the main contribution to the diagram Fig.2 comes 
from regions of internal momenta ${\bf p}_1$ and ${\bf p}_2$ are
restricted to the ${\cal A}$-shell. Inside the ${\cal A}$-shell 
the Green and the D-functions have strong singularities on the 
mass-shell. As it was done in the Appendix B we integrate in 
both short loop the internal frequencies $\omega_1$ and 
$\omega_2$ and find that only non-static term is non-zero:
\bea 
\Sigma_0\sim t^2\int {d^2p_1d^2p_2\over \epsilon^3({\bf p_1})
\epsilon^3({\bf p_2})}{\Omega f^2({\bf p_1,p_2})
f({\bf p_1,p_1})f({\bf p_2,p_2})\over(\Omega-cq\cos\phi_1-ib_1)
(\Omega-cq\cos\phi_2-ib_2)}\times                  \nonumber  \\
{\epsilon^{-2}({\bf p_1-p_2})\over
(\epsilon({\bf p}_1)-\epsilon({\bf p}_2)-
\epsilon({\bf p_1-p_2}))},                               \label{tocy}
\eea
where $\phi_1$ and $\phi_2$ is the direction of the momenta 
${\bf p_1}$ and ${\bf p_2}$. We assumed that $\phi_1\approx\phi_2$.
Since the three spin-wave processes is not allowed by the 
conservation laws, the Green function 
$G(\omega_1-\omega_2,{\bf p_1-p_2})$ is off the mass-shell (the 
corresponding last denominator in Eq.(\ref{tocy}) reads: $(p_1+p_2)
(\phi_1-\phi_2)^2+p_0\cos^2(\phi_1)>0$ (if $p_1,p_2\gg p_0$. Now 
we can count the momenta powers in Eq.(\ref{tocy}) to verify 
that the integration is 
convergent towards the dipolar momentum $p_0$, and, thus has no 
logarithm. A simple counting of temperatures shows that 
Eq.(\ref{tocy}) $\sim t^2/b_1^2\sim t^{1/2}$. Hence the two-loop 
dissipation function is $b_0'=b_0(1+t^{1/4}/L)$. Similar 
consideration shows that the function $b_1'(\psi,q)-b_1(\psi)$, 
which represents the two-loop corrections for $b_1$, is small in 
$t^{1/4}$, and is also small in the ratio $p_0/q$.

\par
Having explicit expressions for the self-energy
we can analyze the dispersion relation 
$\omega^2=\epsilon^2({\bf p})+\Sigma(\omega, {\bf p})$
in the range
of small $\omega$ and $p$. New results are expected
for the region $p<p_c=p_0\sqrt{ t L}$ in which $\Sigma_0$
becomes comparable with $\epsilon^2({\bf p})$.
In a range of momentum $p_{DM}\ll p\ll p_c$
and angles $\psi\ll\sqrt{p_0 t L/p}$, we find a new
propagating soft mode with the dispersion:

\bee
\omega = cp(p^2 + p_0p\psi^2)^{1/2}/p_0\sqrt{ t L}.        \label{soft}
\ene

\par
The dissipation of the soft mode grows to the boundary
of the region and becomes of the order of its energy
at $\psi\sim \sqrt{p_0 t L/p}$ or $p\sim p_{DM}$. There is
no soft mode beyond the indicated range. The spin-wave 
mode persists at $p>p_0 t L$. In a range $p\ll p_{DM}$ and 
small angles a new diffusion mode occurs with the 
dispersion:

\bee
\omega = 
-i\epsilon^2({\bf p})t^{-1/4}L^{5/4}/\beta_0cp_0.        \label{diff}
\ene

\par
The angular range of the diffusion mode increases with
decreasing $p$ and captures the entire circle at
$p<p_0 t L$. 

\par
At the end of this section we would like to remind that equation
(\ref{MR}) was obtained for the quasi-excitation directed along 
the $y$-axis. It can be easily checked that in case of arbitrarily
directed ${\bf q}$ one must write $\sin^2(2\psi+2\phi)$ instead of
$\sin(2\psi)$ in eq. (\ref{B0}) where $\sin\phi=q_x/q$. However, 
in this case the 
integral becomes singular, so one must treat this equation 
more carefully. We will come back to this question in Appendix C.

\section{Renormalization of the diffusion mode.}

\par
In this section we concern ourself with the renormalization of the diffusion
mode. As we established in the previous section, at wavevectors
$p<p_{DM}$ the diffusive dynamics term dominates ($\lambda\rightarrow 0$ 
limit) in the harmonic part of JD Functional (27-28). The interaction
between 'diffusons', given by the anharmonic part of the JDF (32), 
effectively 'renormalizes' the 'diffuson' dispersion at very small 
wavevectors $p<p_a\ll p_{DM}$. We shall determine the anomalous diffusion 
onset wavevector $p_a$ in the end of this section. 

\par
To simplify further calculations, we introduce a scale transformation of
the fields $\phi$, $\phe$
$\rightarrow$ $(Jg/T^2)^{-1/4}{\phi}$, $(Jg/T^2)^{-1/4}{\phe}$. 
In these notations the JDF (28,32) is:
\bea
{\cal J}[{\phi},{\phe}]=\sum_{\omega,{\bf k}}{\phe}_{-\omega,-{\bf k}}
{\Biggl(}a\frac{T}{\Gamma_0}{\phe}_{\omega,{\bf k}}-\left(ak_y^2+
\frac{k_x^2}{a|k_y|}\right){\phi}_{\omega,{\bf k}}-
a\frac{i\omega}{\Gamma_0}{\phi}_{\omega,{\bf k}}            \nonumber    \\
-w\frac{k_{x}k_{y}}{|k_y|}\left[\frac{\phi^2}{2}\right]_{\omega,{\bf k}}
{\!\!\!\!}-
w\sum_{\Omega,{\bf p}}{\phi}_{\omega-\Omega,{\bf k}-{\bf p}}
\frac{p_{x}p_{y}}{|p_y|}{\phi}_{\Omega,{\bf p}}-  
w^{2}a\sum_{\Omega,{\bf p}}{\phi}_{\omega-\Omega,{\bf k}-{\bf p}}|p_y|
\left[\frac{\phi^2}{2}\right]_{\Omega,{\bf p}},
{\Biggr)}
\label{JDDF}
\eea
where $\Gamma_0=J\Gamma$, $a=\sqrt{J/g}$ and 
\bea w=(T^{2}g/J^{3})^{1/4}. \label{wer}
\eea
Note, that one can get rid off the spatial anisotropy charge $a$ 
by rescaling 
the $x$ coordinate only.

\par
Assuming that all terms in the 
r.-h.s. of JDF (\ref{JDDF}) have the same scaling dimensions, equal 
to zero, and accepting the 
dimension of $k_y$ $\Delta_y=1$, $\Delta_a^0=0$ and 
$\Delta_{1/\Gamma}^0=0$, we immediately find from
the second and third terms that $\Delta_{k_x}^0=3/2$. Comparing the second
and the fourth terms, we get $\Delta_{\omega}^0=2$.
The first and the fourth terms give 
$\tilde{\Delta}_{\phi}^0+\Delta_{\omega}^0=\tilde{\Delta}_{\phe}^0$. 
From the 
condition that the scaling dimension of the first term is zero, we have
$2\tilde{\Delta}_{\phe}^0+\Delta_{\omega}^0+1+3/2=0$. 
Finally, comparing the second and the fifth terms, we obtain 
$\Delta_w^0+\Delta_{\omega}^0+\tilde{\Delta}_{\phi}^0+2=0$
Now it is straightforward to find all bare scaling dimensions:
$\Delta_{k_x}^0=3/2$,  
$\Delta_{\omega}^0=2$, $\Delta_{\phi}^0 =1/4$, $\Delta_{\phe}^0=9/4$, 
$\Delta_w^0=1/4$. (Note that the notation 
$\tilde{\Delta}_{{\phi},\phe}^0$ stands for the scaling dimensions of 
$\phi({\bf k},\omega)$ and $\phe({\bf k},\omega)$, 
while $\Delta_{{\phi},\phe}^0$ denotes the scaling dimensions of 
$\phi({\bf x},t)$ and $\phe({\bf x},t)$.)
\par
According to the standard renormalization group procedure, we introduce 
renormalization constants: ${\phi}=Z_{\phi}^{1/2}{\phi}_R$,
${\phe}=Z_{\phe}^{1/2}{\phe}_R$, $a=Z_{a}a_R$,
$1/\Gamma_0=Z_{1/\Gamma}1/\Gamma_R$ and 
$\tilde{w}=Z_{w}\tilde{w}_R$, where 
\bea \tilde{w}=l^{1/4}w. \label{wpow} \eea
First we note 
that the fields $\phi$ and $\phe$ have the same renormalization constants, 
which immediately follows from the fluctuation-dissipation theorem. Indeed,
according to this theorem
$$D (x,y)=\langle\phi (x)\phi (y)\rangle \sim {\cal I}m G (x,y) =  
{\cal I}m \langle\phe (x)\phi (y)\rangle.$$ 

Let us divide the Janssen-De Dominics functional into two parts:
$$ J=J_R+\Delta J.$$ 
The first one is the "renormalized" functional ${\cal J}_R$
\bea
{\cal J}_R[{\phi}_R,{\phe}_R]=\sum_{\omega,{\bf k}}\phe_{R-\omega,-{\bf k}}
{\Biggl(}a_R\frac{T}{\Gamma_R}{\phe}_{R\omega,{\bf k}}-\left(a_Rk_y^2+
\frac{k_x^2}{a_R|k_y|}\right){\phi}_{R\omega,{\bf k}}      \nonumber  \\
-a_R\frac{i\omega}{\Gamma_R}{\phi}_{R\omega,{\bf k}}              
-\tilde{w}_{R}l^{1/4}\frac{k_{x}k_{y}}{|k_y|}
\left[\frac{\phi_{R}^2}{2}\right]_{\omega,{\bf k}}{\!\!\!\!\!\!}-
\tilde{w}_{R}l^{1/4}\sum_{\Omega,{\bf p}}
{\phi}_{R\omega-\Omega,{\bf k}-{\bf p}}
\frac{p_{x}p_{y}}{|p_y|}{\phi}_{R\Omega,{\bf p}}         \nonumber    \\
-\tilde{w}_{R}^{2}a_{R}l^{1/2}\sum_{\Omega,{\bf p}}             
{\phi}_{R\omega-\Omega,{\bf k}-{\bf p}}|p_y|
\left[\frac{\phi_{R}^2}{2}\right]_{\Omega,{\bf p}}
{\Biggr)}.  
\eea
The second one ${\Delta}{\cal J}$ contains the counter-terms: 
\bea
&{\Delta}&{\cal J}=\sum_{\omega,{\bf k}}{\phe}_{R-\omega,-{\bf k}}
{\Biggl(}a_R\frac{T}{\Gamma_R}(Z_{\phi}Z_aZ_{1/\Gamma}-1)
{\phe}_{R\omega,{\bf k}}                                        \nonumber \\  
&-&\left(a_{R}(Z_{\phi}Z_{a}-1)k_y^2+
(Z_{\phi}Z^{-1}_{a}-1)\frac{k_x^2}{a_{R}|k_y|}\right)
{\phi}_{R\omega,{\bf k}}                                         \nonumber\\
&-&a_R\frac{i\omega}{\Gamma_R}(Z_{\phi}Z_aZ_{1/\Gamma}-1)
{\phi}_{R\omega,{\bf k}}                                         
-(Z_{w}Z_{\phi}^{3/2}-1)\tilde{w}_{R}l^{1/4}\frac{k_{x}k_{y}}{|k_y|}     
\left[\frac{\phi_{R}^2}{2}\right]_{\omega,{\bf k}}               \nonumber  \\
&-&(Z_{w}Z_{\phi}^{3/2}-1)\tilde{w}_{R}l^{1/4}\sum_{\Omega,{\bf p}}
{\phi}_{R\omega-\Omega,{\bf k}-{\bf p}}
\frac{p_{x}p_{y}}{|p_y|}{\phi}_{R\Omega,{\bf p}}            \nonumber   \\
&-&(Z_{w}^{2}Z_{a}Z_{\phi}^{2}-1)
\tilde{w}_{R}^{2}a_{R}l^{1/2}         
\sum_{\Omega,{\bf p}}
{\phi}_{R\omega-\Omega,{\bf k}-{\bf p}}|p_y|
\left[\frac{\phi_{R}^2}{2}\right]_{\Omega,{\bf p}}
{\Biggr)}.
\eea

\par
A simple power counting shows that all corrections are only logarithmically
divergent in the dimension of $5/2$ (see section II). Hence, in what 
follows the $\varepsilon$-regularization scheme with $\varepsilon=1/2$ 
is assumed.

\par
Evaluating the diagram shown in Fig.1a up to the second order in $k_y$
(using the bare $G_0$ and $D_0$ functions) 
one finds the one-loop correction to the second term of the JDF:
$$
-\frac{18}{128\pi}a_{R}\tilde{w}^2_{R}
l^{1/2}{\int}\frac{{\di}k_y}{k_y^{3/2}}-
(Z_{\phi}Z_{a}-1)a_{R}.
$$
In order to cancel the divergency, we put:
\bee
Z_{\phi}Z_{a}=1-\frac{18}{128\pi}\tilde{w}^{2}_{R}
l^{1/2}{\int}\frac{\di k_y}{k_y^{3/2}}.               \label{a}
\ene
The same procedure for the fourth term gives:
\bee
Z_aZ_{\phi}Z_{1/\Gamma}=1-\frac1{32\pi}\tilde{w}^{2}_{R}
l^{1/2}{\int}\frac{\di k_y}{k_y^{3/2}}.           \label{ZGamma}
\ene
In the Appendix we show that both the three-leg and four-leg
vertices do not have one-loop corrections. Hence:
\bee
Z_{w}Z_{\phi}^{3/2}=1.                                 \label{w}
\ene
One can easily see that the one-loop correction to the term 
$\phe\frac{k_x^2}{|k_y|}\phi$ in eqn (\ref{w}) vanishes as well (see 
\cite{kashuba,zinn}). 
It means that
\bee
Z_{a}^{-1}Z_{\phi}=1.                                   \label{Zphi}
\ene
Equations (\ref{a},\ref{ZGamma},\ref{w},\ref{Zphi}) have a following 
solution:
\bea
Z_{a}=&Z_{\phi}&=1-\frac{9}{128\pi}
\tilde{w}_{R}^{2}l^{1/2}{\int}\frac{\di k_y}{k_y^{3/2}},      \label{Za}  \\
&Z_{1/\Gamma}&=1+\frac{7}{64\pi}
\tilde{w}_{R}^{2}l^{1/2}{\int}\frac{\di k_y}{k_y^{3/2}},      \label{ZG}  \\
&Z_{w}&=\left(1+\frac{27}{256\pi}
\tilde{w}^{2}_{R}l^{1/2}
{\int}\frac{\di k_y}{k_y^{3/2}}\right).                       \label{Zw}
\eea

Next we introduce the Gell-Mann-Low $\beta$-function 
$\beta={\mu}\left.\frac{\partial w_R}{\partial \mu}\right|_{w,\Lambda}$
and the Callan-Simanzik anomalous dimensions 
$\gamma_{a}=\frac{\mu}{a_R}\frac{\partial a_R}{\partial \mu}$, 
$\gamma_{\Gamma}={\mu}{\Gamma}_R\frac{\partial 1/{\Gamma}_R}{\partial \mu}$ 
and 
$\eta_{\phi}=\frac{\mu}{Z_{\phi}}\frac{\partial Z_{\phi}}{\partial \mu}$.
We denote by $\mu$ the scale at which the coupling constant is equal to
$\tilde{w}_R$
and denote by  $\Lambda$ the scale at which the coupling constant is 
equal to $\tilde{w}$.
From (\ref{Zw}) and the definition of $Z_w$, one has:
\bee
w_R=w\left(\frac{\Lambda}{\mu}\right)^{1/4}
\left(1-\frac{27}{128\pi}\tilde{w}^2{\Lambda}^{1/2}
\int_{\mu}^{\Lambda}\frac{\di k_y}{k_y^{3/2}}\right).
\ene
And finally, we find the $\beta$-function, which coincide with that
found in the statics case \cite{kashuba}:
\bee
\beta(\tilde{w}_R)=-\frac14\tilde{w}_R+\frac{27}{128\pi}\tilde{w}_R^3
\label{fixp} \ene
The fixed point of the renormalization group flow is
$$
\left.\tilde{w}_R^{\ast}\right.^2=\frac{32\pi}{27}
$$
After performing this procedure for anomalous dimensions one gets:
\bea
&\gamma_a&=-\frac{9}{64\pi}\tilde{w}^2_R,                        \\
&\gamma_{1/\Gamma}&=\frac{7}{32\pi}\tilde{w}^2_R,              \\
&\eta_{\phi}&=\frac{9}{64\pi}\tilde{w}^2_R.
\eea
Using the result for $\tilde{w}_R^{\ast}$, it is straightforward to find
${\gamma_a}^{\ast}=-1/6$, ${\gamma_{1/\Gamma}}^{\ast}=7/27$ and
${\eta_{\phi}}^{\ast}=1/6$.
\par
The long range limit of the functions $G$ and $D$ is found from the 
Callan-Symanzik equation \cite{zinn}:
\bea
G(t,{\bf x})=\frac{1}{(x^2+|y|^{8/3})^{7/8}}
f\left(\frac{x}{|y|^{4/3}},\frac{t}{|y|^{47/27}}\right)      \label{sG} \\
D(t,{\bf x})=\frac1{(x^2+|y|^{8/3})^{1/4}}
\tilde{f}\left(\frac{x}{|y|^{4/3}},\frac{t}{|y|^{47/27}}\right)
\eea
where $f(x,y)$ and $\tilde{f}(x,y)$ are arbitrary functions. 
\par
In the static limit ($t=0$) the exponents in the correlation function 
(\ref{sG}) are exact, as was previously found by one of the authors
\cite{kashuba}:
\bee
D({\bf r})\,=\,\langle\phi ({\bf r})\phi (0)\rangle\sim (
 x^{-1/2},\,y^{-2/3})
\label{static}
\ene 

For the Fourier-components of the Green function eq.\ref{sG}, we 
find in the region of anomalous diffuson:
\bee 
G(\omega, {\bf k})\,=\,f_1\left(\frac{k_x}{k_y^{4/3}},\,\frac{\omega}
{k_y^{47/27}}\right) .
\ene
The anomalous dispersion of the diffusion mode announced in the 
Abstract follows from the last equation.
We see that the static dipole
contribution is not renormalized in dynamics, as was  
suggested in (\cite{AhF,Pel}). We also note that the exchange coupling
acquires an anomalous dimension $\Delta_J=1/3$, whereas the dynamic term
$\omega/\Gamma$ acquires anomalous a dimension $+2/27$. Taking
into account the anomalous dimension $\gamma_{1/\Gamma}=7/27$,
we conclude that the anomalous dimension of $\omega$ is
$\gamma_{\omega}=1/6$. The interaction between the
diffusons in the scaling limit reduces the dissipation or,
in other words, hardens the diffuson.

Now we can estimate the wavevector $p_a$, an upper boundary for
anomalous diffusion. We assume that temperature is small (\ref{tcon}). 
Initially, 
according to Eq.\ref{wer}, the bare vertex $w_0=\sqrt{T}g^{1/4}/J^{3/4},$ 
and is also small. Under the renormalization flow, the 
vertex $w_R$ grows with the inverse wavevector as the power $1/4$ 
(\ref{wpow}). The RG flow starts at $p_{DM}$ and approaches 
the fixed point at the root of the Gell-mann-Low function $\beta(w_R)=0$. 
Invoking eq.\ref{fixp}, we
find the fixed point solution $w_R\sim 1$. Thus, the wavevecor 
$p_a$ is defined as the wavevector at which $w_R\sim 1$:  
\bea w_0(p_{DM}/p_a)^{1/4}\sim 1.\eea
We see that
\bea p_a\sim T^2T^{3/4}, \eea
that is very small. Even if $t$ is not small, $p_a\sim p_{DM}w_0^4
\sim p_0(ga/J)$. It is much smaller than $p_0$.

\section{Susceptibilities.}

\par
In this section we find the susceptibility to the magnetic field directed
along the average magnetization $\langle\bf S\rangle$ ($y$-axis), the 
so-called longitudinal susceptibility. We consider
the magnetic field in the form $H=H_{0}+{\delta}H({\bf x},t)$ where $H_0$
is independent of ${\bf x}$ and $t$.
When an additional magnetic field ${\delta}H$ is imposed,
a new vertex ${\delta}h_{\omega,{\bf k}}[{\phi}{\phe}]_{-\omega,-{\bf k}}$
emerges in the JDF (\ref{JDfunc})
(we denote $h=g_{G}{\mu}_{B}SH;\,\,{\delta}h=g_{G}{\mu}_{B}S{\delta}H$). 
It leads to a correction
${\delta}D(\omega,{\bf k})$ to the correlation function 
$D(\omega,{\bf k})$
$$
{\delta}D({\bf x}_{1}={\bf x}_{2},t_{1}=t_{2})=
{\int}_{\!\!\! \Omega}{\int}_{\!\!\! {\bf k}}
{\int}_{\!\!\! \omega}{\int}_{\!\!\! {\bf q}}
h({\omega},{\bf q})D_{0}({\Omega},{\bf k})\left[ 
G_{0}({\Omega}+{\omega},{\bf k}+{\bf q})+
G_{0}^{\ast}({\Omega}-{\omega},{\bf k}-{\bf q})\right],
$$
where $D_{0}(\omega,{\bf k})$ and $G_{0}(\omega,{\bf k})$ are taken from
(\ref{GFunc},\ref{DFunc}).
\par
By definition, the susceptibility $\chi$ is:
\bea
{\chi}({\omega},{\bf k},h)=
\frac{\delta}{{\delta}H({\omega},{\bf k})}{\langle}S_{y}{\rangle}=
-{1 \over 2}\frac{\delta}{{\delta}H({\omega},{\bf k})}
{\langle}{\phi}({\bf x},t){\phi}({\bf x},t){\rangle}       \nonumber    \\
=-\frac{g_{G}{\mu}_{B}S}{2}\frac{\delta}{{\delta}h({\omega},{\bf k})}
D({\bf x}_{1}={\bf x}_{2},t_{1}=t_{2}).                      \label{chi}
\eea
Hence,
\bee
{\chi}({\omega},{\bf q})=
-g_{G}{\mu}_{B}S{\int}_{\!\!\! \Omega}{\int}_{\!\!\! {\bf k}}
D_{0}({\Omega},{\bf k})
G_{0}({\Omega}+{\omega},{\bf k}+{\bf q}).                  \label{chioq}
\ene
In the most interesting case, when ${\bf q}=0$, all integrals can be 
evaluated and the final answer is:
\bea
{\chi}(\omega, h)=g_{G}{\mu}_{B}S\frac{{\Gamma}^{2}(3/4)}{4
{\pi}\sqrt{\pi}}T
\left(\frac{J^{3}}{g^{2}}\right)^{1/4}
\frac{2\Gamma}{\omega}             \nonumber   \\
{\times}\left[ h^{1/4}- \left( h -
\frac{i\omega}{2\Gamma}\right)^{1/4}\right].   \label{chio}
\eea
In the limiting case $\omega=0$, the susceptibility
reads:
 $$\chi = const\cdot h^{-3/4}.$$
This result has been found earlier \cite{maleev-pf}.

\section{Conclusion.}

\par
In conclusion we discuss how new modes can be observed in
experiment. The new modes appear on a macroscopic scale of length
of the order of magnitude $1\mu m$. Even rather weak in-plane
anisotropy can suppress or disguise the new modes. Therefore, 
we can expect that the new dynamics will be observed in films with
very weak in-plane anisotropy. The best known candidates for this 
role are films grown on hexagonal substrates. The hexagonal
anisotropy is naturally weaker than the tetragonal one, because 
they are
proportional to a higher degree of the relativistic parameter.
Besides, the hexagonal anisotropy totally vanishes at large distances 
in a range of 
temperature from $(4/9)T_{BKT}$ till $T_{BKT}$, where 
$T_{BKT}$ is the temperature of Berezinskii-Kosterlitz-Thouless
transition \cite{PU,JKKN}.
The simplest idea is to use the (111) face of FCC crystals, 
such as Ag, Au, Cu. An iron film on the (111) face of Ar 
has been grown by S.Bader and coworkers \cite{bader}. Recently the Ru 
film has been grown on the hexagonal graphite substrate \cite{rau}.
Thus, 2D ferromagnets with exact XY symmetry are available. 

\par
The next question is: what dynamic effects can be observed and at what 
conditions? As we have noted earlier, if the in-plane anisotropy
is not especially small (less than 1K in energy scale), the only 
opportunity is to use a six-fold substrate in the range of temperature,
not small in comparison with $T_{BKT}$. In this situation the 
observation of propagating soft modes described in section III
seems to be improbable, since all they require $t\ll 1$, 
i.e. $T\ll T_c$. However, the anomalous diffusion can be observed
even at $T\sim T_c$, given a sufficiently large scale of 
length ($\geq 10\mu m$). The best way to observe it is to apply
a short and inhomogeneous pulse of magnetic field, and follow
when the secondary signal will arrive to fixed indicators.
Such a picosecond-pulse technique has been recently used
for investigation of the film dynamics \cite{garwin}. We propose to use the
same pulse technique to different films.

\par
A quick estimation of the time $t_{x,y}$ needed for the secondary 
signal to 
reach the indicator at distances $L_x$ and $L_y$ along
${\hat x}$ and ${\hat y}$ directions shows:
$t_{x,y}\sim (\hbar /Jp_0a)(L_{x,y}/a)^{\Delta_{x,y}^a}$,
where $a$ is the lattice constant and $\Delta^a_{x,y}$ are
the anomalous-diffusion dimensions for axes ${\hat x}$ and 
${\hat y}$ respectively. Thus, $t_x\sim\hbar/ga(L_x/a)^{47/36}
\sim 10^{3.5}$sec and $t_y\sim\hbar/ga(L_y/a)^{47/27}
\sim 1$sec., where we have assumed $L_x\sim L_y\sim 1$cm.

\par
The retardation time for the secondary signal is much longer than 
the time for the primary signal propagation. The strong size 
and direction dependence of the propagation time can be used 
for detecting of the anomalous diffusion.

\section{Acknowledgements}

The work of one of the authors (A.K.) was supported in part
by the Swiss National Fond under 'Oststaaten-Sofort\-hilfemassnahmen'
grant N7GUPJ038620. Our thanks are due to M.V. Volpert for her
help in preparation of the manuscript.

\appendix{One-loop vertices corrections}

First we note that the JDF (\ref{JDDF}) has the following 
nontrivial symmetry \cite{kashuba}.

\bea
\phi ({\bf k},\omega) & \rightarrow & \phi ({\bf k},\omega)+\varepsilon 
\delta^2 ({\bf k}) \delta (\omega),             \label{symphi}    \\
\phe({\bf k},\omega) & \rightarrow & \phe({\bf k},\omega),  \label{symphe}  \\
k_x & \rightarrow & k_x - \varepsilon aw k_y,             \label{symkx}    \\
k_y & \rightarrow & k_y.                                  \label{symky}
\eea

The partition function (\ref{Zfunc}) must have the same symmetry
$\delta Z[j,\jh]=0$. By a standard procedure \cite{zinn}, we find
the implications  of 
the symmetry (\ref{symphi}-\ref{symky}), known as Ward-Takahashi 
identities to the so-called "generating functional for proper 
vertices" $\bG [\vphi,\vphe]$ 
(It is the Legendre transform of $\ln Z[j,\jh]$ with respect
to the fields $\vphi$ and $\vphe$):

\bea
\int_{\! \bf k} \int_{\! \omega}\Biggl(\delta (\omega)
\delta ({\bf k})\frac{\delta \bG}{\delta \vphi(-{\bf k},-\omega)}
-aw \biggl[k_y \vphi(-{\bf k},-\omega)\frac{\partial}{\partial k_x}
\frac{\delta \bG}{\delta \vphi (-{\bf k},-\omega)}        \nonumber  \\
+k_y \vphe (-{\bf k},-\omega)\frac{\partial}{\partial k_x}
\frac{\delta \bG}{\delta \vphe (-{\bf k},-\omega)}
\biggl]\Biggl)=0.                                     \label{WT}
\eea

Now, writing down Taylor-like expansion of $\bG[\vphi,\vphe]$
over fields $\vphi$ and $\vphe$ and plugging it into (\ref{WT}),
we find:

\bee
\bG_{\vphi \vphi \vphe}(0,\bar{k},-\bar{k})-
awk_y\frac{\partial}{\partial k_x} 
\bG_{\vphi \vphe}(\bar{k},-\bar{k})=0,               \label{WTT}
\ene
where $\bar{k}$ is used for $({\bf k},\omega)$.
It was proven \cite{zinn} that there is no corrections 
to the term $k_x^2/|k_y|$
due to analyticity. According to (\ref{WTT}), 
it means that corrections 
to the three-leg vertex vanish if we put one of the two frequencies,
which the vertex depends upon, equal to zero. Hence, the corrections must
depend on the product of the two $\omega$s and are small
in the framework of Section III. Considering loop-wise expansion of $\bG$
in the same spirit, one finds:

$$
\bG=\sum_{n=0}^{\infty}w_n \bG^{(n)}\approx\bG^{(0)}+w\bG^{(1)}+\ldots
$$ 
Looking only at the divergent parts of the corrections 
$\bG^{(0)}$ and $\bG^{(1)}$, we see that,
because $\bG^{(0)}$ is just the bare action and has no divergencies at all,
the one-loop corrections to the $\omega$-independent three-leg vertex
do not diverge.
As a consequence of this fact we obtain equation (\ref{w}).

\appendix{Calculation of self-energy}

\newcommand{\ep}{\epsilon({\bf p+q}/2)}
\newcommand{\eps}{\epsilon^2({\bf p+q}/2)}
\newcommand{\epss}{\epsilon^{-2}({\bf p+q}/2)}
\newcommand{\emm}{\epsilon({\bf p-q}/2)}
\newcommand{\ems}{\epsilon^2({\bf p-q}/2)}
\newcommand{\emss}{\epsilon^{-2}({\bf p-q}/2)}
\newcommand{\ff}{f^2({\bf p-q}/2,{\bf p+q}/2)}

Let us start with the Dyson Eq.(42) corresponding to a
one-loop diagram on Fig.1a. We may integrate the Eq.(42) over 
the contour in complex plane $\omega$ such that the poles of 
$G$-function are outside the contour:
\begin{eqnarray}
\Sigma_0={p_0^2c^6t\over2\pi}\int \ff
\left[ \left( (\Omega + \emm )^2 - \eps + ib_1\right)^{-1}
\right.\nonumber\\ \left. +\, \left( (\Omega - \emm)^2 -
\eps - ib_1\right)^{-1} \right] {d^2p\over \ems}.
\end{eqnarray}
It is convinient to change in the second $G$-function ${\bf p}$ on 
${\bf -p}$:
\begin{eqnarray}
\Sigma_0={p_0^2c^6t\over 2\pi}\int {\ff d^2p\over \Omega+\emm-
\ep+ib_1}\nonumber\\ \left[{\emss\over (\Omega+\ep+\emm)}
+{\epss\over(\Omega-\ep-\emm)} \right] 
\end{eqnarray}
Keeping only the lowest order in small momentum ${\bf q}$ and
frequency $\Omega$ and after a simple expansion in the brackets 
we find:
\bea
\Sigma_0= {p_0^2c^6t\over 2\pi}\int {f^2({\bf p,p})d^2p\over\Omega 
+\emm-\ep+ib_1}\times              \nonumber             \\
{-2\Omega+4(\ep-\emm)\over 4\epsilon^4({\bf p})}.
\eea
At this point we separate the above integral into the static
$\Omega$-independent part and the rest 'dynamical' self-energy.
The static $\Omega$-independent self-energy (it always real) 
reads:
\begin{equation} 
\Sigma_{st}=-{p_0^2c^6t\over 2\pi}\int {f^2({\bf p,p})\over
\epsilon^4({\bf p})}d^2p.
\end{equation}
Now let us take into account the static self-energy given by 
the diagram on Fig.1b:
\begin{equation} \Sigma_b={p_0c^4t\over 4\pi}\int {p_y^2-p_x^2\over 
p^2\epsilon^2({\bf p})}d^2p
\end{equation}
Comparing Eq.(B4) (remember that $f({\bf p,p})=p_x^2p_y^2/p^2$) 
and Eq.(B5) we conclude that these cancel each other in the
${\cal A}$-shell. Thus, we have verified explicitly that the
$\Omega,{\bf q}$-independent part of the self-energy is strictly
zero as guaranteed by the Ward identity due to the rotation
symmetry of the system (see Appendix A). To get non-zero static 
self-energy we have to expand self-energies like Eqs.(B4,B5) in 
powers of the transferred momentum ${\bf q}$. This will be done in 
the Section IV using statics field-theoretical technic not dynamical
as in this Appendix. The result is that ${\bf q}$-dependent static
self-energy only matters when the spin-wave momentum ${\bf q}$
is so small that anomalous diffusion sets up. On shorter 
wavelengths like $q<p_{DM}$ we may safely neglect the contribution
of Eq.(B4,B5).

Now we return to the $\Omega$-dipendent dynamical part of the
self energy:
\begin{equation}
\Sigma_0(\chi, q)={p_0^2c^6t\over 2\pi}2\Omega\int {\sin^2(2\phi)
\over\Omega -q\cos\phi +ib_1}{dp\,d\phi\over4c^4 p}
\end{equation}
The integral over $p$ gives exactly the logarithmic factor $L$:
\begin{equation}
\Sigma_0(\chi, q)={p_0^2c^2t\over \pi}L\Omega\int 
{\sin^2\phi\cos^2\phi\over\Omega -q\cos\phi +ib_1}d\phi
\end{equation}
The imaginary part of Eq.(B7) could be easily found in the
limit $\Omega,cq\gg b_1$. In this case we use the formula
$${1\over x-i0}=P{1\over x}+i\pi\delta(x)$$
to find the $\Im$-part of Eq.(46). The $\Re$-part is also
simple to calculate:
\begin{equation}
\Re\Sigma_0(\chi, q)={p_0^2c^2t\over \pi}L{\Omega\over q}\int 
{\sin^2\phi\cos\phi [(-\Omega+q\cos\phi)+\Omega]\over
\Omega -q\cos\phi}d\phi,
\end{equation}
and then simplify:
\begin{equation}
\Re\Sigma_0(\chi, q)={p_0^2c^2t\over \pi}L{\Omega^2\over q}\int 
{\sin^2\phi\cos\phi\over
\Omega -q\cos\phi}d\phi.
\end{equation}
We repeat the same step:
\begin{equation}
\Re\Sigma_0(\chi, q)={p_0^2c^2t\over \pi}L{\Omega^2\over q^2}\int 
{\sin^2\phi(-\Omega+q\cos\phi+\Omega ) \over
\Omega -q\cos\phi}d\phi
\end{equation}
with further simplification:
\begin{equation}
\Re\Sigma_0(\chi, q)={p_0^2c^2t\over \pi}L{\Omega^3\over q^2}\int 
{\sin^2\phi \over \Omega -q\cos\phi }d\phi
 -{p_0^2c^2t\over\pi}L{\Omega^2\over q^2}{2\pi\over 2}
\end{equation}
One could easily continue the same procedure to find
\begin{equation}
\Re\Sigma_0(\chi, q)={p_0^2c^2t}L{\Omega^2\over q^2}(2
{\Omega^2\over q^2}-1)=p_0^2c^2tL\cos^2\chi\cos(2\chi),
\end{equation}
which is exactly the Eq.(46) of the III Section. 

Finally let us show that the self-energy in the ${\cal A}$-shell
is negligible. We first assume that it does negligible: 
$a^2(\omega,{\bf p})\ll \epsilon^2({\bf p})$. Then we consider
the self-energy Eq.(B2), provided the external frequency and 
momentum lies on the mass-shell inside the ${\cal A}$-shell: 
$\Omega=\epsilon({\bf q})$. Let also $\sqrt{\lambda/J}\gg q<p_0$.
One could easily verify that the main contribution comes if
$q\ll p$. In this case we may use the Eq.(B6):
\begin{equation}
\Sigma_0={p_0^2c^6t\over 2\pi}2cq\int {\sin^2(2\phi)
\over c\sqrt{q^2+p_0q_x^2/q}-q\cos\phi +ib_1}{dp\,d\phi\over4c^4 p}
\end{equation}
Integration over $p$ gives $\log(\sqrt{\lambda/J}/q)$, whereas the
integration over the relative direction of internal and external 
spin-waves gives factor $\sqrt{q^3/p_0q^2_x}$. Thus, the real
self-energy $$a^2(\Omega,{\bf q})\sim tLc^2p_0\sqrt{p_0q}.$$
At small $t$ this result justifies our neglecting of self-energy
in the ${\cal A}$-shell.

\appendix{The angular dependence of the self-energy operator.}
\par
In this appendix we analyze the angular dependence of the 
self-energy part found in Section III. Repeating all arguments,
one can find that eq. (\ref{B0}) must be written in 
slightly modified form:

\begin{equation}
\Sigma_0 (\Omega,q,\phi)={c^2p_0^2t\over 4\pi}\int
{c^4p^3dp\over\epsilon^4({\bf p})}{\Omega\sin^2(2\psi+2\phi)
\di \psi\over \Omega-cq\cos\psi+ib_1}.                  \label{B0a}
\end{equation}

Now again, assuming $cq \gg b_1$, one finds:

\bee
\Sigma_0 (r,\phi)=c^2 p_0^2 tL \left[r^2 \left(
2r^2-1-2r\sqrt{r^2-1}\right)\cos(4\phi)+\frac12\frac{r}{\sqrt{r^2-1}} 
\sin^2(2\phi)\right],                                  \label{B00a}
\ene
where $r=\Omega/cq$. In the case $\phi=0$ and 
$r=\cos \chi$ the formula (\ref{B00}) is recovered.
\par
The main contribution to the integral in eq. (\ref{sss})
comes from a region of very small angles $\phi$. Hence, the
correction to the $\Sigma_0$ we have found above is not important
in this calculations and equation (\ref{B1}) still holds.
\par
By the next step we need to plug $b_1 (\psi)$ from (\ref{B1}) 
to the nonzero-angle form of equation (\ref{MR})
 
\bee 
\Sigma (\Omega,q,\phi)=
i\Omega c^2 p_0^2 tL\int_0^{2\pi}\frac{\di\psi}{4\pi}
\frac{\sin^2(2\psi+2\phi)}{ib_1(\psi)+
\Omega-cq\cos\psi}.                                    \label{MRa} 
\ene
Evaluating this integral, one finds:

\bea
\Sigma(|\Omega| \ll cq) & = & i\Omega c\tilde{p}
\left[\cos^2(2\phi)-\alpha\left(\frac{q}{p_{DM}}\right)^{-3/5}
\sin^2(2\phi)\right]                             \label{sig1}  \\
\Sigma(|\Omega| \gg cq) & = & i\Omega c\tilde{p}
\left[\cos^2(2\phi)-\alpha\left(\frac{|\Omega|}{cp_{DM}}\right)^{-3/5}
\sin^2(2\phi)\right],                            \label{sig2}  
\eea
where we denote $\tilde{p}=\beta_0 t^{1/4}L^{5/4}p_0$,
$p_{DM}=\beta_1 t^{3/4}L^{-1/4}p_0$ and 
$\alpha=2^{-1/5}/\beta_0 \beta_1 \approx 0.54$. 
In what follows we will also use the notation $p_c=p_0\sqrt{tL}$.
While equation (\ref{B00a}) for $\Sigma$ holds
for $cq,\Omega \gg b_1 (\psi)$, equations (\ref{sig1}) 
and (\ref{sig2}) are valid in the opposite case $cq, \Omega \ll b_1(\psi)$.
\par
Now we can analyze the dispersion relation  
$\omega^2 ({\bf q})=\epsilon^2({\bf q})+\Sigma (\omega, {\bf q})$ more
accurately. In experiment usually $t \ll 1$ so $p_0 \gg \tilde{p} \gg p_c \gg
p_{DM}$. Easy, but tedious calculations show that there can exist
up to 9 
asymptotic regions in the momentum space with different 
dispersion relations. 

\bea
\begin{array}{lll}
 \omega^2 =c^2p_0 q \sin^2\phi,
& \mbox{if}& 
p_{DM}\frac{\sin^{3/2}(2\phi)\sin(\phi/2)}{\cos\phi}\ll      
q \ll p_0 \sin^2\phi
\\
 \omega^2 =\frac12 c^2 q^3 \frac{p_0}{p_c^2}\phi^2,
&\mbox{if}&
\frac{p_c^2}{p_0}\phi^2 
\left(1+\frac{p_0 p_{DM}}{p_c^2}\phi^{1/2}\right)
\ll q \ll \frac{p_c^2 p_0 \phi^2}{p_0^2\phi^4+p_c^2}         \label{reg2}     
\\
 \omega^2 =\frac12 c^2 q^3 \frac{p_0}{p_c^2},
&\mbox{if}&
p_{DM}(\pi/2-\phi)^{1/2} \ll q \ll \frac{p_c^2}{p_0}         \label{reg3}
\\
 \omega^2 =\frac{c^2}{2}\frac{q^4}{p_c^2},
&\mbox{if}&
p_0 \phi^2 \ll q \ll p_c                                     \label{reg4}     
\\
 \omega =-icq^2 \frac{p_0}{p_c^2}, 
&\mbox{if}&
p_{DM}\phi^{5/2} \ll q \ll \frac{p_c^2}{p_0}\phi^2           \label{reg5}     
\\
 \omega =-icq\frac{p_0}{\tilde{p}}\phi^2,
&\mbox{if} & 
p_{DM}\phi^{10/3}\ll q\ll p_{DM}\phi^{5/2};\,\,\,\,\,\phi\ll
\sqrt{\tilde{p}/p_0}                                        \label{reg6}
\\
 \omega =-icq\frac{p_0}{\tilde{p}}\phi^2,
&\mbox{if} & 
p_{DM}\phi^{4/3}\ll q\ll p_{DM}\phi^{1/2};\,\,\,\,\,\
\sqrt{\tilde{p}/p_0}\ll \phi\ll 1                           \label{reg7}    
\\
 \omega =-icq\frac{p_0}{\tilde{p}},
&\mbox{if}&
\frac{\tilde{p} p_{DM}}{p_0}(\pi/2-\phi)^{10/3} \ll q
\ll \frac{\tilde{p} p_{DM}}{p_0}(\pi/2-\phi)^{1/2}          \label{reg8}
\\
 \omega^{2/5} =-iq\frac{c^{2/5}p_0}{4\alpha\tilde{p}p_{DM}^{3/5}}
\frac1{\cos^2\phi},
&\mbox{if} & 
p_{DM}({\tilde{p}\over p_0})^{5/3}(\pi/2-\phi)^{10/3}\ll q\ll 
\frac{\tilde{p}p_{DM}}{p_0}\phi^{4/3}(\pi /2-\phi)^{10/3}         \label{reg9} 
\end{array}                                                       \nonumber
\eea
where $0< \phi < \pi/2$

\par
In figure 3 we show
the case $t\sim 0.3$, $L\sim 1$ and $p_0=1$.

{\Large\bf Figure Caption}
\begin{itemize}
\item FIG.1. a,b) The main contribution to the self-energy. The functions
$G$ and $D$ are given by (\ref{GMatrixTot}). The three-leg vertices in
fig.1.a and the four-leg vertex in fig.1.b are from (\ref{Jinteraction}).
c) Two-loop correction to the self-energy. Momenta of internal lines are
indicated.
\item FIG.2. The most 'dangerous' two-loop diagram.
\item FIG.3. Maps of the regions with different dispersion relations. 
\end{itemize}

\begin{thebibliography}{22}
\bibitem{constants} The classical exchange constant $J$ and anisotropy
$\lambda$ used in this paper differs from that in quantum Heisenberg model:
$J_{\mbox{cl}}=J_{\mbox{qua}}S^2$ and ${\lambda}_{\mbox{cl}}=
{\lambda}_{\mbox{qua}}S^2$
\bibitem{Rfunction} An allowed by symmetry term  
${\sim}R'{\dot{\cal \pi}}{\dot{\cal \pi}}$ in the dissipative function
is of higher order in time derivatives, since $\dot{\cal \pi}\sim 
\ddot{\phi}$.
dissipation functional are also irrelevant.
\end{thebibliography}
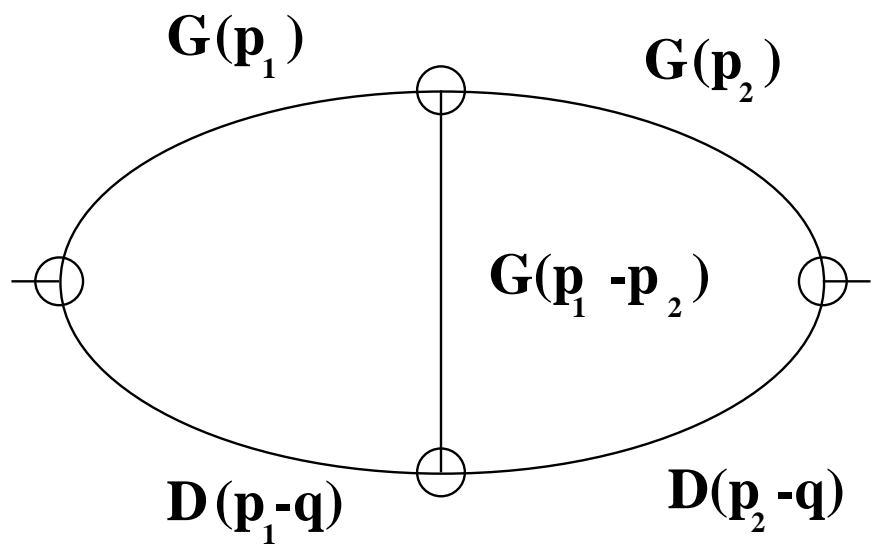
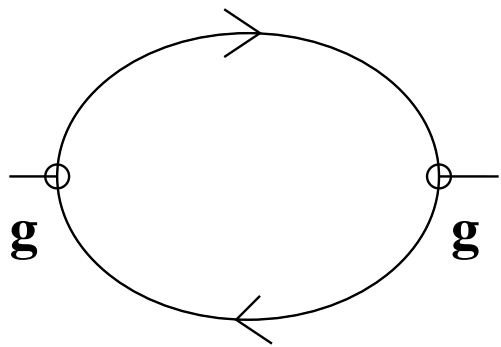
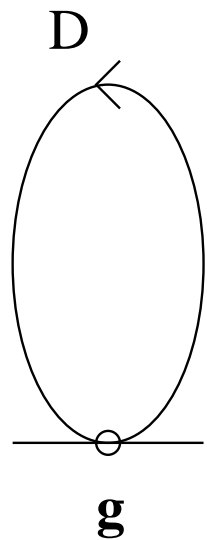
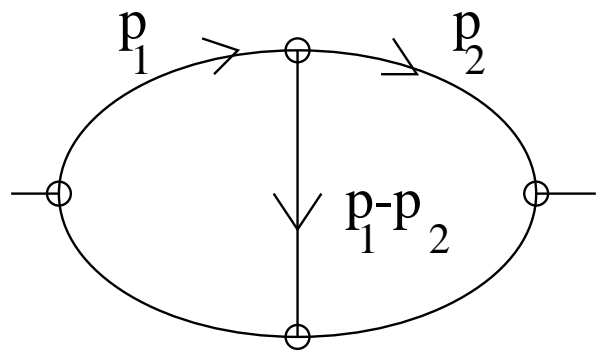
\end{document}